# A partial differential equation for pseudocontact shift

G.T.P. Charnock[1], Ilya Kuprov[2,*]

[1]*Oxford e-Research Centre, University of Oxford, 7 Keble Road, Oxford OX1 3QG, UK.*

[2]*School of Chemistry, University of Southampton, Highfield Campus, Southampton, SO17 1BJ, UK.*

[*]Corresponding author (i.kuprov@soton.ac.uk )

**Abstract**

It is demonstrated that pseudocontact shift (PCS), viewed as a scalar or a tensor field in three dimensions, obeys an elliptic partial differential equation with a source term that depends on the Hessian of the unpaired electron probability density. The equation enables straightforward PCS prediction as well as analysis of experimental PCS data in systems with multiple and / or distributed unpaired electron centres.

**Introduction**

Pseudocontact shift (PCS) is an effective additional contribution to the nuclear chemical shift that arises in open-shell chemical systems due to partial polarization of the electron spin by the applied magnetic field [1]. The primary application of PCS is in structural biology, where it provides additional distance restraints for molecular structure determination [2,3]. *Pseudocontact* shift is different from the *contact* shift in that it does not require electron-nuclear overlap and propagates instead through the dipolar coupling [1].

It can be verified by direct inspection that the commonly used point electron dipole expression for the PCS [4] has a zero Laplacian everywhere except the origin:

$$\sigma_{\text{PCS}}^{(\text{point})} = \frac{1}{12\pi}\left[\Delta\chi_{\text{ax}}\frac{2z^2-x^2-y^2}{r^5}+\frac{3}{2}\Delta\chi_{\text{rh}}\frac{x^2-y^2}{r^5}\right]; \qquad \nabla^2\sigma_{\text{PCS}}^{(\text{point})}=0 \qquad (1)$$

where $\Delta\chi_{\text{ax}}$ and $\Delta\chi_{\text{rh}}$ are axiality and rhombicity of the electron magnetic susceptibility tensor $\chi$ and the nucleus is located at $[x,y,z]$ relative to the electron. This is to be expected – all classical electromagnetic phenomena must obey Maxwell's equations – but the singularity at the origin also suggests the possibility of an elliptic partial differential equation existing for the harder case of the PCS generated by a *non-point* electron probability density $\rho(\vec{r})$:

$$\sigma_{\text{PCS}}^{(\text{non-point})} = \int \sigma_{\text{PCS}}^{(\text{point})}(\vec{r}-\vec{r}')\rho(\vec{r}')d^3\vec{r}'; \qquad \nabla^2\sigma_{\text{PCS}}^{(\text{non-point})}=\kappa(\chi,\vec{r}) \qquad (2)$$

in which the source term $\kappa(\chi,\vec{r})$ is unknown and has so far resisted all derivation attempts: a direct calculation of the Laplacian of the convolution of Equation (1) with a finite unpaired electron density comes across singular integrals that cannot be regularized [5]. Yet the prize is tempting – elliptic partial differential equations are a classical topic in mathematics and a simple enough PDE would generalize all PCS analysis problems, improve data interpretation close to the unpaired electron location and also provide a way of measuring spin label probability distributions in biomolecular EPR experiments [6]. It would be convenient too – numerical PDE solvers have been available in standard software packages for a long time. In this communication we derive the equation and comment on some of its properties.

**Hyperfine shift as a total energy derivative**

To facilitate subsequent mathematics, and also for the sake of completeness, we provide in this section succinct derivations, using the relatively modern total energy derivative formalism [7], of the classical expressions for the various components of the hyperfine shift tensor [1,4]. For his-torical reasons we shall separate the *hyperfine shift tensor* $\boldsymbol{\sigma}_{HF}$ into the *contact shift tensor* $\boldsymbol{\sigma}_{CS}$ and the *dipolar shift tensor* $\boldsymbol{\sigma}_{DS}$. Their isotropic parts shall be called *contact shift* and *pseudocontact shift*, and denoted $\sigma_{CS}$ and $\sigma_{PCS}$ respectively.

Placed in a magnetic field $\vec{B}_0$, a point electron at the origin with a magnetic susceptibility tensor $\chi \ll 1$ would acquire an average magnetic moment:

$$\vec{\mu}_e = \boldsymbol{\chi} \cdot \vec{B}_0 / \mu_0 \tag{3}$$

The additional magnetic field $\vec{B}_1$ generated by this dipole at the point $\vec{r}$ is:

$$\vec{B}_1 = \frac{\mu_0}{4\pi}\left[3\frac{\vec{r} \otimes \vec{r}^{\,\mathrm{T}}}{r^5} - \frac{1}{r^3}\right] \cdot \vec{\mu}_e = \frac{1}{4\pi}\left[3\frac{\vec{r} \otimes \vec{r}^{\,\mathrm{T}}}{r^5} - \frac{1}{r^3}\right] \cdot \boldsymbol{\chi} \cdot \vec{B}_0 \tag{4}$$

The change in energy $E$ for a nuclear magnetic dipole $\vec{\mu}_N$ located at that point would be:

$$E = -\vec{\mu}_N^{\mathrm{T}} \cdot \vec{B}_1 = -\frac{1}{4\pi}\vec{\mu}_N^{\mathrm{T}} \cdot \left[3\frac{\vec{r} \otimes \vec{r}^{\,\mathrm{T}}}{r^5} - \frac{1}{r^3}\right] \cdot \boldsymbol{\chi} \cdot \vec{B}_0 \tag{5}$$

and therefore, the additional chemical shift tensor experienced by the nucleus, measured relative to the unperturbed conditions, would be:

$$\boldsymbol{\sigma}_{DS} = -\frac{\partial^2 E}{\partial \vec{\mu}_N^{\mathrm{T}} \partial \vec{B}_0} = \frac{1}{4\pi}\left[3\frac{\vec{r} \otimes \vec{r}^{\,\mathrm{T}}}{r^5} - \frac{1}{r^3}\right] \cdot \boldsymbol{\chi} \tag{6}$$

where the minus appears because of the relationship between chemical shielding and chemical shift [8,9]. The isotropic part of this tensor is:

$$\sigma_{PCS} = \frac{1}{3}\mathrm{Tr}(\boldsymbol{\sigma}) = \frac{1}{12\pi}\mathrm{Tr}\left(\left[3\frac{\vec{r} \otimes \vec{r}^{\,\mathrm{T}}}{r^5} - \frac{1}{r^3}\right] \cdot \boldsymbol{\chi}\right) \tag{7}$$

For a point electron located at $\vec{r}_e$ and a point nucleus located at $\vec{r}_N$ the final expression is:

$$\sigma_{PCS}(\vec{r}_N, \vec{r}_e) = \frac{1}{12\pi}\mathrm{Tr}\left(\left[3\frac{(\vec{r}_N - \vec{r}_e) \otimes (\vec{r}_N - \vec{r}_e)^{\mathrm{T}}}{|\vec{r}_N - \vec{r}_e|^5} - \frac{1}{|\vec{r}_N - \vec{r}_e|^3}\right] \cdot \boldsymbol{\chi}\right) \tag{8}$$

A neat derivation for the hyperfine shift can also be given in terms of the hyperfine coupling tensor $\mathbf{A}$. The spin Hamiltonian for the hyperfine interaction is $\hat{H} = \hat{\vec{I}} \cdot \mathbf{A} \cdot \hat{\vec{S}}$ and therefore the corresponding energy change for the $k$-th nucleus in the system is:

$$E = \frac{\vec{\mu}_n^{(k)} \cdot \mathbf{A}^{(k)} \cdot \vec{\mu}_e}{\gamma_e \gamma_n^{(k)} \hbar} = \frac{\vec{\mu}_n^{(k)} \cdot \mathbf{A}^{(k)} \cdot \boldsymbol{\chi} \cdot \vec{B}_0}{\mu_0 \gamma_e \gamma_n^{(k)} \hbar} \tag{9}$$

After using the same derivative expression for the chemical shielding [7], we obtain:

$$\boldsymbol{\sigma}_{CS}^{(k)} + \boldsymbol{\sigma}_{DS}^{(k)} = -\frac{\mathbf{A}^{(k)} \cdot \boldsymbol{\chi}}{\mu_0 \gamma_e \gamma_n^{(k)} \hbar}, \quad \sigma_{CS}^{(k)} + \sigma_{PCS}^{(k)} = -\frac{1}{3}\mathrm{Tr}\left[\frac{\mathbf{A}^{(k)} \cdot \boldsymbol{\chi}}{\mu_0 \gamma_e \gamma_n^{(k)} \hbar}\right] \tag{10}$$

A significant advantage of these equations over other physically equivalent formulations is that the excitation structure of the quantum chemistry part of the problem is hidden from the user – both the hyperfine tensor and the susceptibility tensor are effective quantities that already incorporate all of the formidable complexities of the electronic structure theory [7]. The derivations given above are classical, but the only assumption in Equation (10) is that $\chi \ll 1$ (meaning also that magnetic hyperpolarizability terms in the electronic structure theory are negligible) and that the electron relaxes sufficiently rapidly for Equation (3) to always remain valid.

Equation (8), in its various forms, has done considerable service to the NMR community over the last forty years – naturally occurring calcium, magnesium and other metals in biological systems can often be substituted with lanthanides and pseudocontact shift then used to obtain distance restraints for structure determination purposes [2,3]. The point electron dipole assumption does, however, have its validity range – Equation (8) is not applicable in close proximity of the metal centre, most notably in lanthanide spin labels, and also in the cases where the electron probability density is broadly distributed within the molecular structure.

**Derivation of the elliptic PDE**

The source term $\kappa(\chi,\vec{r})$ in Equation (2) for pseudocontact shift induced by a distributed electron is not currently known. The most straightforward derivation is to notice that:

$$3\frac{(\vec{r}_N - \vec{r}_e)\otimes(\vec{r}_N - \vec{r}_e)^T}{|\vec{r}_N - \vec{r}_e|^5} - \frac{1}{|\vec{r}_N - \vec{r}_e|^3} = \frac{\partial^2}{\partial \vec{r}_e \partial \vec{r}_e^T}\frac{1}{|\vec{r}_N - \vec{r}_e|} \tag{11}$$

and to note that the Laplacian of the reciprocal distance is:

$$\nabla_N^2 \frac{1}{|\vec{r}_N - \vec{r}_e|} = -4\pi\delta^3(\vec{r}_N - \vec{r}_e) \tag{12}$$

With these observations in place, we can conclude that:

$$\nabla_N^2 \sigma_{PCS}(\vec{r}_N, \vec{r}_e) = -\frac{1}{3}\mathrm{Tr}\left(\left[\frac{\partial^2}{\partial \vec{r}_e \partial \vec{r}_e^T}\delta^3(\vec{r}_N - \vec{r}_e)\right]\cdot \chi\right) \tag{13}$$

The convolution of this expression with a finite electron probability density distribution $\rho(\vec{r}_e)$ then yields (after dropping the N subscript on the nuclear coordinate vector $\vec{r}_N$):

$$\nabla^2 \sigma_{PCS}(\vec{r}) = -\frac{1}{3}\mathrm{Tr}\left(\left[\frac{\partial^2 \rho(\vec{r})}{\partial \vec{r} \partial \vec{r}^T}\right]\cdot \chi\right) \tag{14}$$

After abbreviating the Hessian of $\rho(\vec{r})$ in square brackets

$$\mathbf{H}_\rho = \frac{\partial^2 \rho(\vec{r})}{\partial \vec{r} \partial \vec{r}^{\mathrm{T}}} \qquad (15)$$

we arrive at a neat final result:

$$\nabla^2 \sigma_{\mathrm{PCS}} = -(1/3)\mathrm{Tr}\left[\mathbf{H}_\rho \cdot \boldsymbol{\chi}\right] \qquad (16)$$

A similar derivation for the full 3×3 dipolar shift tensor in Equation (6) yields:

$$\nabla^2 \boldsymbol{\sigma}_{\mathrm{DS}} = -\mathbf{H}_\rho \cdot \boldsymbol{\chi} \qquad (17)$$

If chemical *shielding* is considered instead of the chemical *shift*, the minus sign on the right hand side disappears. Simplicity of Equations (16) and (17) stands in sharp contrast with the unfathomable spherical harmonic expansions [10] generated by *ab initio* treatments using the ligand field theory.

**Analytical and numerical solutions**

General solution strategies for Equations (16) and (17) are identical to those of Poisson's equation [11]. The special case of the point electron in Equation (7) is recovered by observing that the Green's function of the Laplacian is:

$$\nabla^2 G(\vec{r},\vec{r}') = \delta^3(\vec{r}-\vec{r}') \quad \Rightarrow \quad G(\vec{r},\vec{r}') = -\frac{1}{4\pi}\frac{1}{|\vec{r}-\vec{r}'|} \qquad (18)$$

and taking its convolution with the right hand side of Equation (16) for the point electron:

$$\sigma_{\mathrm{PCS}} = \frac{1}{12\pi}\int \frac{1}{|\vec{r}-\vec{r}'|}\mathrm{Tr}\left[\frac{\partial^2}{\partial \vec{r}'^{\mathrm{T}} \partial \vec{r}'}\delta^3(\vec{r}')\boldsymbol{\chi}\right]d^3\vec{r}' = \frac{1}{12\pi}\mathrm{Tr}\left[\frac{\partial^2}{\partial \vec{r}^{\mathrm{T}} \partial \vec{r}}\frac{1}{|\vec{r}|}\boldsymbol{\chi}\right] =$$
$$= \frac{1}{12\pi}\mathrm{Tr}\left(\left[3\frac{\vec{r}\otimes\vec{r}^{\mathrm{T}}}{r^5} - \frac{1}{r^3}\right]\cdot\boldsymbol{\chi}\right) \qquad (19)$$

Analytical treatments of more general unpaired electron probability density distributions may be simplified significantly by noting that the Laplacian in Equations (16) and (17) leaves spherical harmonics intact. General solutions of those equations therefore simply inherit the multipolar expansion from the source term in the same way as Poisson's equation solutions do [12]:

$$\sigma(r,\theta,\varphi) = \frac{1}{4\pi}\sum_{l=0}^{\infty}\sum_{m=0}^{l}(2-\delta_{m0})\frac{(l-m)!}{(l+m)!}\left(\frac{1}{r}\right)^{l+1}P_l^m(\cos\theta)\left(a_{lm}\cos m\varphi + b_{lm}\sin m\varphi\right)$$
$$a_{lm} + ib_{lm} = \int_{\mathbb{R}^3}\kappa(\boldsymbol{\chi},\vec{r})r^l P_l^m(\cos\theta)e^{im\varphi}dV \qquad (20)$$

where $\kappa(\boldsymbol{\chi},\vec{r})$ is either the right hand side of Equation (16), or each of the nine matrix elements on the right hand side of Equation (17).

This is the general *analytical* solution for Equation (16) and for each matrix element of Equation (17), but in practice the interpretation of multipole moments with $l > 2$ is difficult. We would argue here that *numerical* treatment is easier computationally and also more interpretable from the physical point of view because the fundamental quantity there is the unpaired electron probability density $\rho(\vec{r})$. Sparse matrix representations of 3D Laplacians with Dirichlet, Neumann or periodic boundary conditions are readily available [13], and the numerical solution to Equations (16) and (17) requires a single sparse matrix-inverse-times-vector operation, for example:

$$\sigma_{\text{PCS}} = -(1/3)\mathbf{L}^{-1}\text{vec}\left[\text{Tr}\left(\mathbf{H}_\rho \cdot \boldsymbol{\chi}\right)\right] \tag{21}$$

where $\mathbf{L}$ is a matrix representation of the 3D Laplacian with appropriate boundary conditions [13] and $\text{vec}\left[\text{Tr}\left(\mathbf{H}_\rho \cdot \boldsymbol{\chi}\right)\right]$ denotes the index transformation that stretches $\text{Tr}\left(\mathbf{H}_\rho \cdot \boldsymbol{\chi}\right)$, which is a three-dimensional array, into a column vector. Equation (21) has been implemented into the latest version of our *Spinach* library [14]. On a contemporary workstation using Matlab the example solution given in Figure 1 takes a few minutes to compute. The conclusion from Figure 1 is that, for lanthanide-containing spin labels, the accuracy of the point PCS model is very high, presumably due to the localized nature of the *f* orbitals, and it is mostly the contact shift that is making their interpretation difficult in practice.

**Inverse problems**

The most interesting possibility offered by Equations (16) and (17) is the model-free recovery of the electron probability density distribution $\rho(\vec{r})$ and the susceptibility tensor $\boldsymbol{\chi}$ from nuclear coordinates and pseudocontact shifts. The problem may be formulated as a minimization condition, with respect to $\vec{\rho} = \text{vec}\left[\rho(\vec{r})\right]$ and $\boldsymbol{\chi}$, for the following least squares error functional:

$$\begin{aligned}\Omega(\vec{\rho}) &= \left\|\mathbf{P}\mathbf{L}^{-1}\mathbf{K}\vec{\rho} - \vec{\sigma}_{\text{expt}}\right\|^2 + \lambda_1\langle\vec{\rho}|\ln(\vec{\rho})\rangle + \lambda_2\|\mathbf{L}\vec{\rho}\|^2 \\ \mathbf{K}\vec{\rho} &= -(1/3)\text{vec}\left[\text{Tr}\left(\mathbf{H}_\rho \cdot \boldsymbol{\chi}\right)\right]\end{aligned} \tag{22}$$

where $\mathbf{P}$ is the matrix that projects out pseudocontact shifts at the nuclear locations, $\vec{\sigma}_{\text{expt}}$ is a vector of experimental PCS data and $\lambda_{1,2}$ are Tikhonov regularization parameters [15]. The uncommon choice of regularization operators (maximum entropy *and* minimum Laplacian norm) is dictated by two practical considerations:

1. Electron probability distribution is expected to be localized in at least one dimension – even extended conjugated systems, such as porphyrin and carotene radicals, have electron spin densities that closely follow the bonding network. Maximum entropy regularization is known to favour highly local solutions unencumbered by baseline noise [16], hence the presence of $\lambda_1 \langle \vec{\rho} | \ln(\vec{\rho}) \rangle$ term in Equation (22).

2. Electron probability distribution is not infinitely sharp – some penalty should be placed that allows a measure of broadening. Because electron spin densities often have symmetric distributions in PCS systems, high multipoles should also be discouraged in the solution. Both objectives are accomplished by $\lambda_2 \|\mathbf{L}\vec{\rho}\|^2$ term in Equation (22).

$\Omega(\vec{\rho})$ is non-linear with respect to $\vec{\rho}$, necessitating the use of numerical minimization methods (LBFGS [17] is used here), but good initial guesses may be obtained by noting that, for $\lambda_1 = 0$, the global minimum of $\Omega(\vec{\rho})$ with respect to $\vec{\rho}$ is analytical:

$$\vec{\rho}_{\min} = \left(\mathbf{A}^\mathrm{T}\mathbf{A} + \lambda_2 \mathbf{L}^\mathrm{T}\mathbf{L}\right)^{-1} \mathbf{A}^\mathrm{T} \vec{\sigma}_{\mathrm{expt}}, \qquad \mathbf{A} = \mathbf{P}\mathbf{L}^{-1}\mathbf{K} \qquad (23)$$

A synthetic example of computing the PCS field generated by multiple paramagnetic centres and then recovering their distribution from PCS data is given in Figure 2. Additional constraints on the probability density are non-negativity, fixed integral and zero boundary conditions:

$$\rho(\vec{r}) \geq 0 \qquad \int \rho(\vec{r}) d^3\vec{r} = N \qquad \rho(\infty) = \rho'(\infty) = 0 \qquad (24)$$

where $N$ is the number of unpaired electrons in the system. Because the error functional in Equation (22) has two regularization parameters, a generalization of the *L*-curve method [18] to surfaces [19] is used here. Better regularization methods for Equation (16) that could improve the fidelity of the reconstruction in Figure 2C are undoubtedly possible, but are beyond the scope of the present work.

**Conclusions**

Equations (16) and (17) provide a simple and numerically friendly alternative to voluminous and abstruse multipolar expansions in situations where electron probability distributions deviate significantly from the point electron case. Both forward (PCS from $\rho(\vec{r})$ and $\chi$) and backward ($\chi$ and $\rho(\vec{r})$ from PCS) calculations are straightforward – the former is accomplished in a single sparse matrix-inverse-times-vector operation prescribed by Equation (21) and the latter is an instance of Tikhonov regularization of the well-researched source recovery problem for an elliptic PDE [20] with the error functional specified in Equation (22).

Attention should also be drawn again to the simple general connection between hyperfine shift and hyperfine coupling provided by Equations (10) – modern electronic structure theory packages are able to compute both hyperfine tensors and magnetic susceptibility tensors, meaning that hyperfine shift tensors may be obtained essentially for free after standard magnetic property runs in ADF [21], ORCA [22] or Gaussian [23], subject only to the electron spin relaxation time being sufficiently short for the approximation made in Equation (3) to be valid.

The source code for all examples provided above, as well as numerical infrastructure functions (3D finite difference operators, 3D interpolation operators, Tikhonov solvers, volumetric scalar field visualizer, *etc.*), are available in versions 1.5 and higher of *Spinach* library [14].

**Acknowledgements**

The authors are grateful to Alan Kenwright, Claudio Luchinat, Gottfried Otting, Giacomo Parigi, David Parker and Guido Pintacuda for stimulating discussions. This work is supported by EPSRC (EP/H003789/1).

**References**

[1] I. Bertini, C. Luchinat, G. Parigi, *Magnetic susceptibility in paramagnetic NMR*, Progress in Nuclear Magnetic Resonance Spectroscopy 40 (**2002**) 249-273.

[2] I. Bertini, C. Luchinat, G. Parigi, *Paramagnetic constraints: An aid for quick solution structure determination of paramagnetic metalloproteins*, Concepts in Magnetic Resonance 14 (**2002**) 259-286.

[3] G. Otting, *Protein NMR Using Paramagnetic Ions*, Annual Review of Biophysics 39 (**2010**) 387-405.

[4] I. Bertini, C. Luchinat, G. Parigi, Solution NMR of paramagnetic molecules: applications to metallobiomolecules and models, Elsevier, 2001.

[5] G.T.P. Charnock, Computational Spin Dynamics and Visualisation of Large Spin Systems, in: DPhil thesis, Department of Computer Science, University of Oxford, 2012.

[6] G. Jeschke, V. Chechik, P. Ionita, A. Godt, H. Zimmermann, J. Banham, C.R. Timmel, D. Hilger, H. Jung, *DeerAnalysis2006 - a comprehensive software package for analyzing pulsed ELDOR data*, Applied Magnetic Resonance 30 (**2006**) 473-498.

[7] M. Kaupp, M. Bühl, V.G. Malkin, Calculation of NMR and EPR parameters: theory and applications, Wiley, 2004.

[8] R.K. Harris, *Conventions for tensor quantities used in NMR, NQR and ESR*, Solid State Nuclear Magnetic Resonance 10 (**1998**) 177-178.

[9] C.J. Jameson, *Reply to "Conventions for tensor quantities used in nuclear magnetic resonance, nuclear quadrupole resonance and electron spin resonance spectroscopy"*. Solid State Nuclear Magnetic Resonance 11 (**1998**) 265-268.

[10] R.M. Golding, L.C. Stubbs, *The theory of the pseudocontact dontribution to NMR shifts of $d^1$-transition and $d^2$-transition metal-ion systems in sites of octahedral symmetry*, Journal of Magnetic Resonance 40 (**1980**) 115-133.

[11] S.D. Poisson, *Remarques sur l'équation qui se présente dans la théorie des attractions des sphéroïdes*, Nouveau Bulletin des Sciences par la Société Philomatique de Paris 3 (**1812-1813**) 388-392.

[12] N. Takaaki, A. Shigeru, *A projective method for an inverse source problem of the Poisson equation*, Inverse Problems 19 (**2003**) 355.

[13] A. Knyazev, M. Argentati, I. Lashuk, E. Ovtchinnikov, *Block Locally Optimal Preconditioned Eigenvalue Xolvers (BLOPEX) in Hypre and PETSc*, SIAM Journal on Scientific Computing 29 (**2007**) 2224-2239.

[14] H.J. Hogben, M. Krzystyniak, G.T.P. Charnock, P.J. Hore, I. Kuprov, *Spinach - A software library for simulation of spin dynamics in large spin systems*, Journal of Magnetic Resonance 208 (**2011**) 179-194.

[15] A.N. Tikhonov, Numerical methods for the solution of ill-posed problems, Kluwer, 1995.

[16] P. Hodgkinson, H.R. Mott, P.C. Driscoll, J.A. Jones, P.J. Hore, *Application of maximum entropy methods to 3-dimensional NMR spectroscopy*, Journal of Magnetic Resonance 101 (**1993**) 218-222.

[17] D.C. Liu, J. Nocedal, *On the Limited Memory BFGS Method for Large-Scale Optimization*, Mathematical Programming 45 (**1989**) 503-528.

[18] P. Hansen, D. O'Leary, *The Use of the L-Curve in the Regularization of Discrete Ill-Posed Problems*, SIAM Journal on Scientific Computing 14 (**1993**) 1487-1503.

[19] B. Murat, E.K. Misha, L.M. Eric, *Efficient determination of multiple regularization parameters in a generalized L-curve framework*, Inverse Problems 18 (**2002**) 1161.

[20] I. Kazufumi, L. Ji-Chuan, *Recovery of inclusions in 2D and 3D domains for Poisson's equation*, Inverse Problems 29 (**2013**) 075005.

[21] G. te Velde, F.M. Bickelhaupt, E.J. Baerends, C.F. Guerra, S.J.A. Van Gisbergen, J.G. Snijders, T. Ziegler, *Chemistry with ADF*, Journal of Computational Chemistry 22 (**2001**) 931-967.

[22] F. Neese, *The ORCA program system*, Computational Molecular Science 2 (**2012**) 73-78.

[23] M.J. Frisch, G.W. Trucks, H.B. Schlegel, G.E. Scuseria, M.A. Robb, J.R. Cheeseman, G. Scalmani, V. Barone, B. Mennucci, G.A. Petersson, H. Nakatsuji, M. Caricato, X. Li, H.P. Hratchian, A.F. Izmaylov, J. Bloino, G. Zheng, J.L. Sonnenberg, M. Hada, M. Ehara, K. Toyota, R. Fukuda, J. Hasegawa, M. Ishida, T. Nakajima, Y. Honda, O. Kitao, H. Nakai, T. Vreven, J.A. Montgomery Jr., J.E. Peralta, F. Ogliaro, M.J. Bearpark, J. Heyd, E.N. Brothers, K.N. Kudin, V.N. Staroverov, R. Kobayashi, J. Normand, K. Raghavachari, A.P. Rendell, J.C. Burant, S.S. Iyengar, J. Tomasi, M. Cossi, N. Rega, N.J. Millam, M. Klene, J.E. Knox, J.B. Cross, V. Bakken, C. Adamo, J. Jaramillo, R. Gomperts, R.E. Stratmann, O. Yazyev, A.J. Austin, R. Cammi, C. Pomelli, J.W. Ochterski, R.L. Martin, K. Morokuma, V.G. Zakrzewski, G.A. Voth, P. Salvador, J.J. Dannenberg, S. Dapprich, A.D. Daniels, Ö. Farkas, J.B. Foresman, J.V. Ortiz, J. Cioslowski, D.J. Fox, Gaussian 09, in, Gaussian, Inc., Wallingford, CT, USA, 2009.

[24] L.S. Natrajan, N.M. Khoabane, B.L. Dadds, C.A. Muryn, R.G. Pritchard, S.L. Heath, A.M. Kenwright, I. Kuprov, S. Faulkner, *Probing the Structure, Conformation, and Stereochemical Exchange in a Family of Lanthanide Complexes Derived from Tetrapyridyl-Appended Cyclen*, Inorganic Chemistry 49 (**2010**) 7700-7709.

[25] A.D. Becke, *A New Mixing of Hartree-Fock and Local Density-Functional Theories*, Journal of Chemical Physics 98 (**1993**) 1372-1377.

[26] C.T. Lee, W.T. Yang, R.G. Parr, *Development of the Colle-Salvetti Correlation-Energy Formula into a Functional of the Electron-Density*, Physical Review B 37 (**1988**) 785-789.

[27] R.A. Kendall, T.H. Dunning, R.J. Harrison, *Electron affinities of the first-row atoms revisited. Systematic basis sets and wave functions*, The Journal of Chemical Physics 96 (**1992**) 6796-6806.

[28] M. Kaupp, P.V. Schleyer, H. Stoll, H. Preuss, *Pseudopotential Approaches to Ca, Sr, and Ba Hydrides - Why Are Some Alkaline-Earth MX$_2$ Compounds Bent*, Journal of Chemical Physics 94 (**1991**) 1360-1366.

[29] T.A. Keith, R.F.W. Bader, *Calculation of magnetic response properties using a continuous set of gauge transformations*, Chemical Physics Letters 210 (**1993**) 223-231.

**Figure captions**

**Figure 1** An example of the forward problem solution and a demonstration of the mutual consistency of Equations (8), (10) and (17). **(A)** Point model *versus* Equation (16) for the PCS in the complex of europium(III) with 1,4,7,10-tetrakis(2-pyridylmethyl)-1,4,7,10-tetraazacyclododecane [24] using the magnetic susceptibility tensor obtained from a DFT calculation. **(B)** Point model *versus* the PCS part of Equation (10) using hyperfine tensors and the magnetic susceptibility tensor from a DFT calculation. **(C)** Volumetric stereo plot of the PCS field computed using Equation (16) with the electron probability density and the susceptibility tensor obtained from a DFT calculation. **(D)** Stereo plot of electron spin density isosurface (at the isovalues of ±0.0004) in the complex. In all cases, the molecular geometry was optimized and the electron spin density estimated using DFT UB3LYP method [25,26] in vacuum with cc-pVTZ basis set [27] on light atoms and Stuttgart ECP basis set on europium [28]. CSGT DFT UB3LYP [29] method with the same combination of basis sets was used to estimate the magnetic susceptibility tensor. The points refer to the symmetry-unique atoms in the ligand, excluding nitrogens that have significant contact shifts. Simulation source code is available within the example set of *Spinach* library version 1.5 and later [14].

**Figure 2** An example of the inverse problem solution where the electron probability distribution is recovered from pseudocontact shift data. **(A)** Volumetric stereo plot of a model system with three electrons with a randomly assigned susceptibility tensor and Gaussian probability distributions randomly positioned within a 20x20x20 Angstrom cube. **(B)** Volumetric stereo plot of the pseudocontact shift field obtained from the probability density cube shown in (A) using Equation (21). **(C)** Volumetric stereo plot of the electron probability distribution obtained by solving the inverse problem as described in the main text. Pseudocontact shift was sampled at 500 random points emulating nuclear locations within the volume and fed into Equation (22), which was then minimized from a random initial guess. Simulation source code is available within the example set of *Spinach* library version 1.5 and later [14].

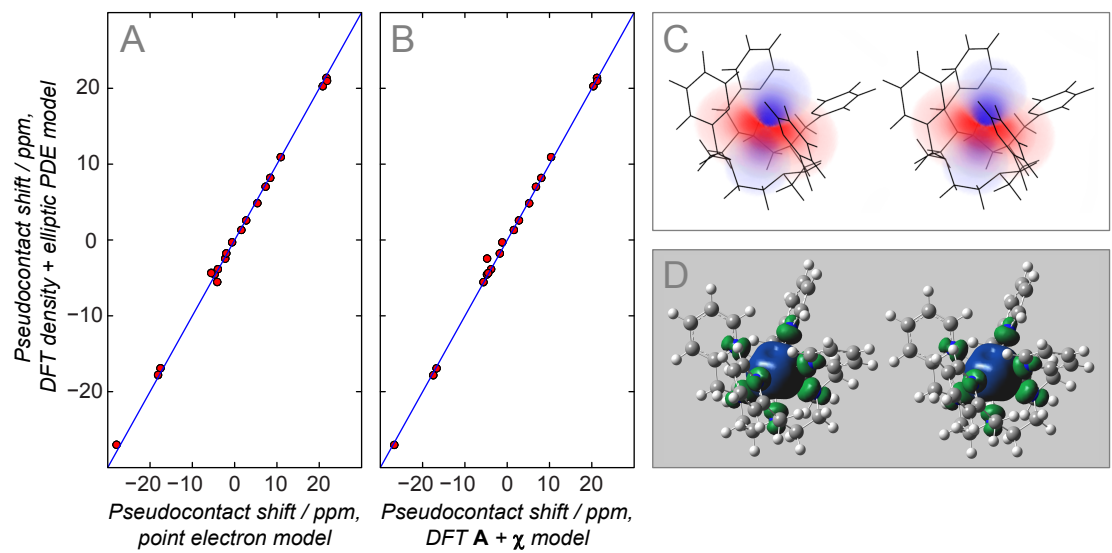

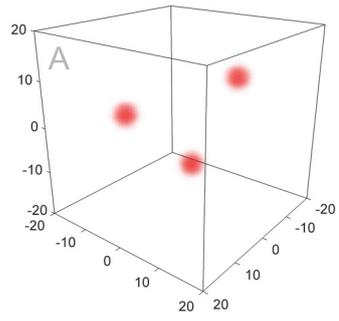 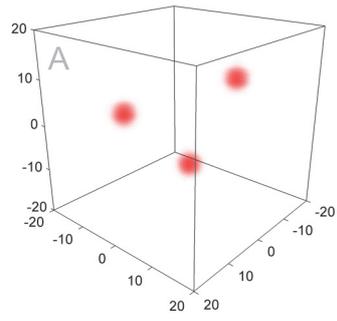

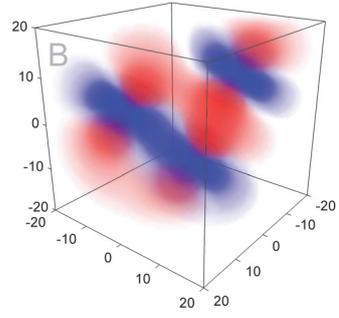 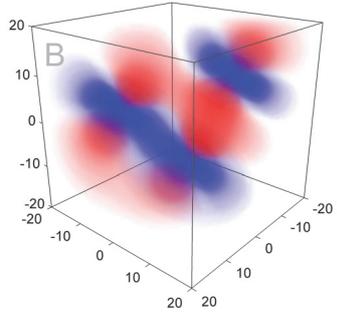

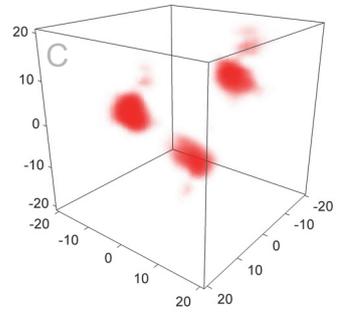 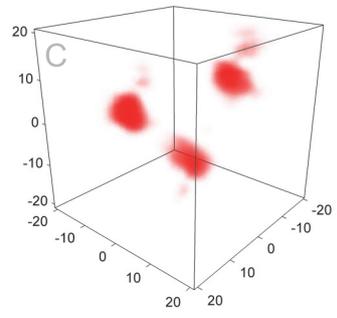